\documentclass[runningheads]{llncs}
\usepackage{fullpage}
\usepackage[T1]{fontenc}
\usepackage{graphicx}
\usepackage{amsmath}
\usepackage{algorithm}
\usepackage{algorithmic}
\usepackage{enumitem} 
\usepackage{xcolor}
\usepackage{comment}

\newtheorem{definition1}{Lemma}
\newtheorem{definition2}{Proposition}
\newtheorem{definition3}{Theorem}
\newtheorem{definition4}{Corollary}

\definecolor{darkgreen}{rgb}{0.0,0.5,0.0}
\definecolor{brown}{rgb}{0.65,0.16,0.16}
\usepackage[utf8]{inputenc}
\usepackage{graphicx} 
\usepackage{subcaption} 

\begin{document}
	
	\title{Dichotomy study of the Steiner tree problem in split-like graphs}
	\titlerunning{Dichotomy study of the Steiner tree problem in split-like graphs}
	
	\author{Jyothish S\inst{1} \and
		N Sadagopan\inst{1} 
	}
	\authorrunning{Jyothish S et al.}
	
	\institute{Indian Institute of Information Technology, Design and Manufacturing,
		Kancheepuram, Chennai, India \\
		\email{\{cs24d0012,sadagopan\}@iiitdm.ac.in}}

	\maketitle
	
	\begin{abstract}

		Given a connected graph $G$ and a terminal set $R \subseteq V(G)$, the minimum Steiner tree problem (ST) asks for a tree that spans all of $R$ with at most $r$ vertices from $V(G)\backslash R$, for some integer $r\geq 0$. A \emph{split graph} is a graph which can be partitioned into a clique and an independent set. It is known from (Garey et al.,1977 \cite{garey}) that ST is NP-complete, even for split graphs \cite{white}. We introduce the class of \emph{split-like graphs} which unifies several known graph classes like bipartite graphs, split graphs, and bisplit graphs \cite{inbook}, allowing for a cohesive study across multiple structural constraints. 
		
		\setlength{\parindent}{2em} We investigate the computational complexity of the Steiner tree problem under structural constraints, specifically  $K_{1,r}$-free, bounded diameter, chordality and star-convexity. Through reductions (primarily from Exact-3-Cover and its variants), the paper establishes a series of dichotomy results.
		
		It precisely gives the boundary for $K_{1,r}$-free bipartite graphs: ST is in P for $r \le 3$ and NP-complete for $r \ge 4$; whereas on $K_{1,r}$-free bisplit graphs, ST is in P for any fixed $r\geq 3$. On bisplit graphs, the Steiner tree problem admits a polynomial-time solution when the diameter is 2. In contrast, for diameters 3 and 4, the problem is NP-complete. The problem is NP-complete under star convexity constraints on the independent set. When star convexity is imposed on the 
		$k$-clique side, the problem is solvable in polynomial time. The problem is NP-complete on chordal bipartite graphs and chordal split graphs (i.e., split graphs themselves), while polynomial-time algorithms exist for other subclasses of split-like graphs.

		\renewcommand{\thefootnote}{\fnsymbol{footnote}}
		\footnotetext{This work is partially supported by NBHM-02011/24/2023/6051 and ANRF (DST)-CRG/2023/007127}
	\end{abstract}
	\section{Introduction}
	
	The Steiner tree problem is one of the most important combinatorial optimization problems, which can be used as a model in many fields, such as global routing, network routing, VLSI design, optical and wireless communication systems, transportation and distribution networks, and phylogenetic tree reconstruction. It is one among 21 NP-complete problems listed by Karp \cite{Karp1972}.The Steiner tree problem is a common term for a class of combinatorial optimization problems defined in various settings. Generally speaking, this problem requires an optimum interconnect for a given set of points under a predefined objective function.
	
	Given a connected edge-weighted graph \(G\) and a subset of vertices \(R \subseteq V(G)\) called terminals, the objective is to find a tree \(T\) that spans all vertices in \(R\) while minimizing the total weight of the edges in \(T\). In the unweighted setting, the goal becomes minimizing either the number of edges in the Steiner tree or the number of additional vertices \(Q \subseteq V(G) \setminus R\) (called Steiner vertices) used to connect the terminals. ST generalizes two well-known problems: the Minimum Spanning Tree when \(R = V(G)\), and the Shortest Path problem when \(|R| = 2\). In this paper, we shall work with the unweighted version of the problem. All graphs in this paper are assumed to be simple unweighted connected unless explicitly mentioned otherwise. 
	
	Since ST is a hard problem for general graphs, it has been studied in different graph classes as well as for approximation results. ST remains NP- complete even on graph classes like Planar \cite{GareyJohn}, Bipartite \cite{Karp1972}, Split \cite{white} and Chordal bipartite graphs \cite{MULLER1987257}. While the problem is solvable in polynomial time for certain graph classes, such as Series Parallel \cite{Seriespar}, Strongly Chordal \cite{white}, Permutation \cite{COLBOURN1990179} and Circle graph \cite{DEFIGUEIREDO2022184}.
	
	A Split graph is a graph that can be partitioned into a clique and an independent set. A \emph{star} $K_{1,r}$ is a complete bipartite graph with one of the partition contains only one vertex and other contains $r$ vertices. ST in $K_{1,4}$-free split graphs is polynomial-time solvable and NP-complete in $K_{1,5}$-free split graphs \cite{RENJITH2020246}. We are trying to extend this dichotomy to more general graph class namely split-like graphs. We define a \emph{split-like graph} as a graph $G$ which can be partitioned into two sets $K$ and $I$, where $I$ is an independent set and the induced subgraph of $K$ is a complete graph or complete $k$-partite graph, for some $k\geq 1$. For each $k = 1, 2, \dots$, the instances are distinct; one is not contained in another. Split-like graph is \emph{perfect}, which is a popular graph class, for every induced sub graph, chromatic number is same as clique number.
	
	We study the classical complexity of ST on split-like graphs subject to four structural parameters: the exclusion of the star  $K_{1,r}$ as a forbidden induced subgraph for some $r\geq 3$, a bounded diameter $d\geq 2$, chordality and star-convexity. The dichotomy results are presented explicitly in Table \ref{tab:dichotomy1}, Table \ref{tab:dichotomy2} and Table \ref{tab:dichotomy3} .

	\begin{figure}
		\centering
		\includegraphics[width=.9\linewidth]{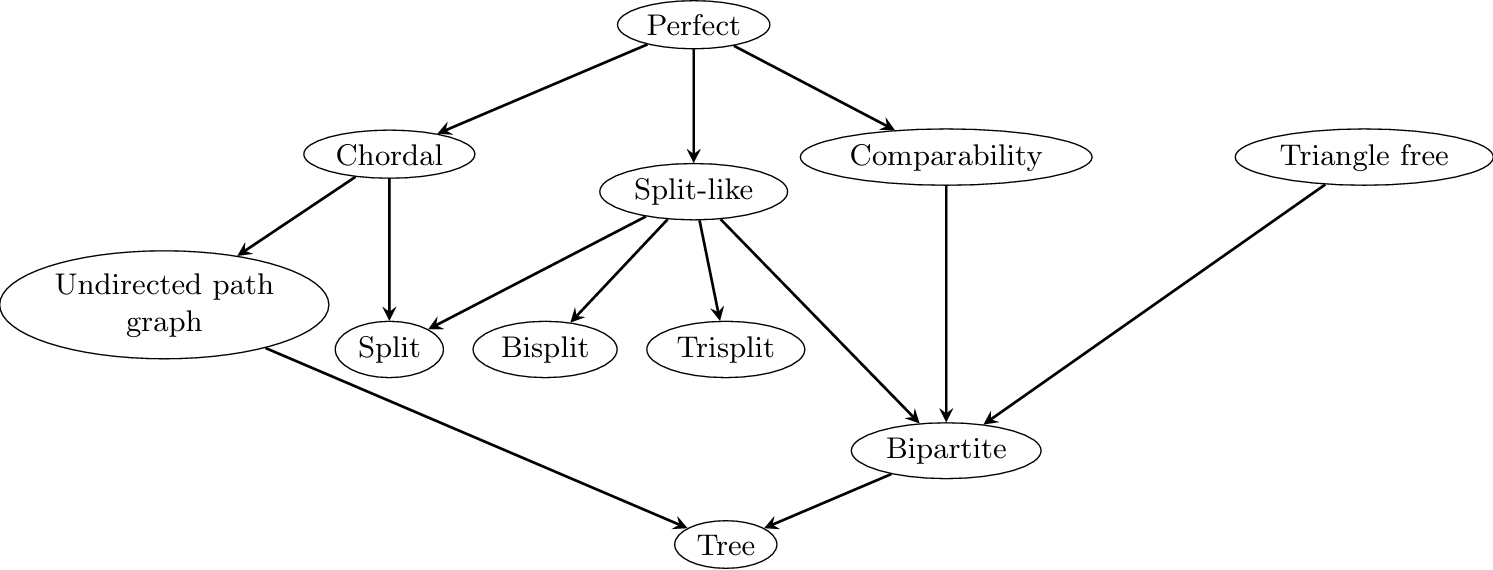}
		\caption{Containment relation of graph classes}
		\label{fig:graph_class}
	\end{figure}

	\begin{table}[h]\label{T1}
		\centering
		\caption{Hardness of ST for $K_{1,r}$-free graphs. \textcolor{brown}{Brown}- Restriction trivializes the problem or Hardness inherited from subclass or tractability from super class, \textcolor{darkgreen}{Green}- Open, \textcolor{blue}{Blue}- From literature, \textcolor{red}{Red}- Our result.}
		\label{tab:dichotomy1}
		\begin{tabular}{|c|c|c|c|c|c|c|}
			\hline
			& Chordal& Triangle free& Bipartite & Split  & Bisplit  & Trisplit \\
			& & & (T\ref{thm2}) & \cite{RENJITH2020246}  & (T\ref{thm3})  & (T\ref{thm10})\\ \hline
			$r=3$ & \textcolor{darkgreen}{Open} &  \textcolor{blue}{NPC}\cite{article3}&\textcolor{red}{P} & \textcolor{blue}{P} & \textcolor{red}{P} & \textcolor{red}{P} \\ \hline
			$r=4$ & \textcolor{red}{NPC} (C\ref{thm6})&\textcolor{brown}{NPC}& \textcolor{red}{NPC} & \textcolor{blue}{P} & \textcolor{red}{P} & \textcolor{red}{P} \\ \hline
			$r=5$ & \textcolor{brown}{NPC} &\textcolor{brown}{NPC}& \textcolor{red}{NPC} & \textcolor{blue}{NPC} & \textcolor{red}{P} & \textcolor{red}{P} \\ \hline
		\end{tabular}
	\end{table}
	\begin{table}[h]\label{T2}
		\centering
		\caption{Hardness of ST for graphs with diameter $d$ as parameter.}
		\label{tab:dichotomy2}
		\begin{tabular}{|c|c|c|c|c|c|}
			\hline
			& Bipartite& Split& Bisplit &Bipartite Bisplit& Trisplit   \\ 
			& & &  &\cite{inbook}&   \\ \hline
			$d=2$ &\textcolor{brown}{P} & \textcolor{brown}{P}&\textcolor{brown}{P} & \textcolor{brown}{P}& \textcolor{brown}{P}  \\ \hline
			$d=3$ & \textcolor{darkgreen}{Open} &\textcolor{blue}{NPC}\cite{white}&\textcolor{red}{NPC} (P\ref{pro6}) & \textcolor{blue}{P}& \textcolor{red}{NPC}(P\ref{protri})  \\ \hline
			$d=4$ & \textcolor{blue}{NPC}\cite{Karp1972} &-& \textcolor{brown}{NPC} &\textcolor{blue}{NPC}& \textcolor{red}{NPC}(P\ref{protri1})  \\ \hline
			$d=5$ & \textcolor{darkgreen}{Open} &-& - &-& -  \\ \hline
		\end{tabular}
	\end{table}
	\begin{table}[h]\label{T3}
		\centering
		\caption{Hardness of ST for graphs with chordality and star convexity as parameter.}
		\label{tab:dichotomy3}
		\begin{tabular}{|c|c|c|c|c|c|}
			\hline
			& Bipartite& Split& Bisplit &Bipartite Bisplit& Trisplit   \\ \hline
			chordal &\textcolor{blue}{NPC}\cite{MULLER1987257} & \textcolor{brown}{NPC}&\textcolor{blue}{P}\cite{inbook} & \textcolor{blue}{P}\cite{inbook}& \textcolor{red}{P}(T\ref{Thr9})  \\ \hline
			star convex & \textcolor{red}{NPC}(T\ref{Th13}) &\textcolor{blue}{NPC}\cite{JOUR}&\textcolor{red}{NPC} (T\ref{A4}) & \textcolor{darkgreen}{Open}& \textcolor{brown}{NPC}  \\ 
			(I/K)&  &\textcolor{blue}{P}&\textcolor{red}{P} (T\ref{T21}) & \textcolor{brown}{P}& \textcolor{brown}{P} \\ \hline
			
		\end{tabular}
	\end{table}

	\subsection{Preliminaries}
	
	In this paper, we work with connected, simple, unweighted graphs. For a graph $G$, the vertex set is $V(G)$ and the edge set is $E(G) = \{(u,v) | u,v \in V(G) $ and $u$ is adjacent to $v \in G$ and $u\neq v\}$. The neighborhood of a vertex $v$ denoted by $N_G(v)=\{u|\{u,v\}\in E(G)\}$. The degree of a vertex $v$ is $deg_G(v)=|N_G(v)|$. For a graph $G$ and $S\subseteq V(G)$, $G[S]$ represents the subgraph of $G$ induced on the vertex set $S$. The diameter of a graph is the farthest distance between any two of its vertices.
	
	A \emph{bipartite} graph $G=A\cup B$ is such that $V(G)$ can be partitioned into two independent sets $A$ and $B$. $K_{m,n}$ is a \emph{complete bipartite} graph with $|A|=m$ and $|B|=n$. A \emph{split} graph $G=K+ I$ is such that $V(G)$ can be partitioned into a clique $K$ and an independent set $I$.  A \emph{bisplit}  graph $G=[A\cup B]+ I$ is such that $V(G)$ can be partitioned into three independent sets $A$,$B$ and $I$, such that $G[A\cup B]$ is a complete bipartite graph.
	
	We define \emph{split-like graph} as a graph $G$ which can be partitioned into two sets $K$ and $I$, where $I$ is an independent set and the induced subgraph $G[K]$ is a complete graph or complete $k$-partite graph, for some $k\geq 1$. For a complete graph or $k=1,2$, $G$ is a split graph, a bipartite graph and a bisplit graph, respectively. For all other values of $k$, the graph is referred to as a $k$-split graph (in particular, when $k=3$, we refer to the graph as a \emph{Trisplit graph} for convenience).
	
	A $ H$-free graph is a graph that does not contain $H$ as an induced subgraph. A graph is \emph{chordal} if every cycle of length greater than 3 has a chord. A graph \( G \) is a \emph{chordal bipartite} graph if \( G \) is bipartite and every cycle of length greater than 4 has a chord.
	
	A split-like graph $G(K,I)$ is called a \emph{$\pi$-convex graph with convexity on $K$}
	if there exists an associated structure $\pi = (K, F)$ such that for each $v \in I$, 
	its neighborhood $N^K_G(v)$ induces a connected subgraph in $\pi$. Similarly, we define \emph{$\pi$-convex graph with convexity on $I$}. $\pi$ can be structures like \emph{tree}, \emph{star}, \emph{path}, \emph{cycle}, etc.\\
	
	We now formally define the minimum Steiner tree problem.\\
	\noindent\fbox{%
		\parbox{\textwidth}{%
			Optimization problem- ST (G, R):
			\\Input: Graph $G$, terminal set $R \subseteq V(G),$
			\\Question: Find a Steiner tree connecting $R$ with minimum number of Steiner vertices $S$.
		}%
	}
	
	\noindent\fbox{%
		\parbox{\textwidth}{%
			Decision problem- ST(G,R,$k$):
			\\Input: Given a graph $G$, terminal set $R \subseteq V(G)$ and an integer $k$
			\\Question: Is there a Steiner tree consisting of at most $k$ Steiner vertices?
		}%
	}
	\section{Steiner tree problem in split-like graphs }
	\subsection{$K_{1,r}$-free split-like graphs}
	ST in $K_{1,4}$-free split graphs is polynomial-time solvable and NP-complete in $K_{1,5}$-free split graphs\cite{RENJITH2020246}. We are trying to extend this study to other split-like graphs. First we are restricting our study to $K_{1,r}$-free bipartite graphs, for a fixed $r \geq 3$. Our dichotomy result establishes that the presence of the claw $K_{1,3}$ makes the ST computationally hard in bipartite graphs. We start with bipartite graphs, then bisplit and extend to k-split graphs.
	\begin{definition1}
		If $G$ is a $K_{1,3}$- free bipartite graph, then $G$ is either an even cycle or a path.
	\end{definition1}
	
	ST in cycles and paths are polynomial-time solvable, which gives the following result.
	\begin{definition2}\label{prp1}
		ST in a $K_{1,3}$- free bipartite graph is polynomial time solvable. 
	\end{definition2}
	
	We now consider the following constrained variant of the Exact-3-Cover problem, which serves as the basis for our reduction showing that ST in a $K_{1,5}$- free bipartite graph is NP-Complete. \\
	
	\noindent\fbox{%
		\parbox{\textwidth}{%
			$EXACT-3-COVER(X,C)-3$\\
			Instance: A collection C of 3-element subsets of a set $X = \{x_1,x_2,...,x_{3q}\}$ such that no element occurs in more than three subsets. \\ 
			Question: Is there a sub-collection $C' \subseteq C$ such that for every $x_i \in X$,$x_i$  belongs to exactly one member of $C'$?
		}%
	}
	\\
	This problem, abbreviated as $X3C-3$, is known to be NP-complete \cite{garey}.\\
	\begin{definition2}
		ST on $K_{1,5}$- free bipartite graphs is NP-complete.
	\end{definition2}
	\begin{proof}
		\textbf{ST is in NP :} Given a certificate S, we show that there exists a deterministic polynomial-time algorithm for verifying the validity of the certificate S. Note that the standard Breadth First Search algorithm can be used to check whether $S \cup R$ is connected. It is easy to check whether $|S| \leq k$. Therefore, the certificate verification can be done in $O(|V(G)| + |E(G)|)$. Thus, we conclude that the Steiner tree problem is in NP.\\
		\textbf{ST is NP-Hard:} An instance of $X3C$-$3(X,C)$ is reduced to an instance of $ST(G,R,k)$ as follows:
		$A=X,B=\{v_i|c_i\in C\}$ . $E(G(A \cup B))=\{(v_i,u) | v_i\in B ,u\in A \text{ and }  u\in c_i\}$. Now, create a binary tree with $B$ as leaf nodes: pair two vertices in $B$ and introduce a common neighbour. If  $|B|$ is odd, the remaining vertex is made adjacent to a new node. Repeat this process to the new set of nodes until they become connected. Let $T$ be the new set of nodes (Figure~\ref{fig:ST1}).\\
		\textbf{$G(A\cup B\cup T)$ is a  $K_{1,5}-$ free connected bipartite graph:} $G(B\cup T)$ is a binary tree. Therefore, it is a $K_{1,4}$- free connected bipartite graph. Now we introduce $A$ (an independent set) and corresponding edges to $G(B\cup T)$, which is again a connected bipartite graph, because each $u\in A$ is adjacent to at least one $v_i\in B $. Since each $u\in B$ is adjacent to exactly three $v_i\in A $, the maximum degree of a vertex in $B$ is 4. Also, the degree of a vertex in $A$ is at most 3. Thus $G(A\cup B\cup T)$ is a  $K_{1,5}$- free connected bipartite graph.\\
		\begin{figure}[h]
			\centering
			\includegraphics[width=0.35\linewidth]{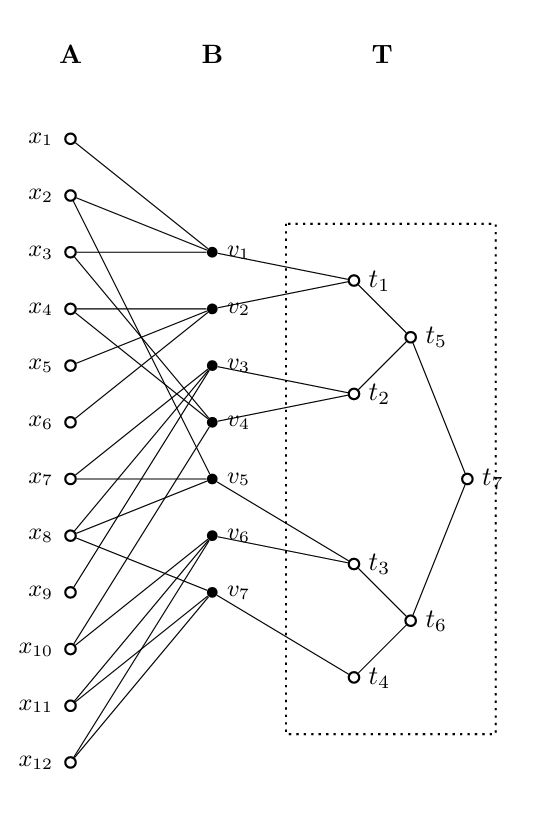}
			\caption{Reduction: An instance of $X3C-3$ to $ST$ in $K_{1,5}$- free bipartite graph}
			\label{fig:ST1}
		\end{figure}
		\textbf{Claim}: $X3C-3(X,C)$ if and only if $ST(G,R,k)$.\\
		\textbf{Proof}.\textit{Necessity}: If there exists $C'\subseteq C$,$|C'|=\frac{|X|}{3}$ which covers all the elements of $X$, then the set of vertices $S=\{v_i \in B|c_i \in C' \} $ forms a Steiner set in $G$ with $R=A\cup T$, where $|S|=\frac{|X|}{3}$.\\
		\textit{Sufficiency}: If there exists a Steiner set $S\subseteq B$ in $G$ on at most $k=\frac{|X|}{3}$ Steiner vertices, then $\forall v\in S, d^{A}(v)=3, |S|=\frac{|X|}{3} \text{ and } N^{A}(S)=X$, which implies that there does not exist $u,v \in S$ such that 	$N^{A}(u)\cap N^{A}(v)\neq \emptyset$. Therefore the set $C'=\{c_i\in C|v_i \in S \}$ forms an exact-3-cover of $X$.
	\end{proof}
	
	Particularly, ST is NP-complete for a $K_{1,5}$- free bipartite graph $G$ with $S \subseteq \{v\in G: deg(v)=4\}$. Can be named as ST-P.\\
	Now we reduce this particular version of ST on $K_{1,5}$- free bipartite graphs to ST on $K_{1,4}$- free bipartite graphs. 
	
	\begin{definition2}\label{A1}
		ST on $K_{1,4}$- free bipartite graphs is NP-complete.
	\end{definition2}
	\begin{proof}
		An instance of $ST-P(G,R,k)$ is reduced to an instance of \\  
		$ST(G',R',k')$ as follows: Consider the graph $G$. Let $v \in G$ be a vertex of degree 4. We split $v$ into two vertices $v_1$ and $v_2$ of degree 2. Then add one extra vertex $u$ and edges $(v_1,u), (w,v_2)$ .\\
		\textbf{$G'$ is a  $K_{1,4}$- free connected bipartite graph:}\\
		Since each degree 4 vertex is replaced with a $P_3$ by splitting the neighbours, $G'$ is again connected bipartite, and the maximum degree becomes 3 (Figure~\ref{fig:ST2}).  
		\begin{figure}[h]
			\centering
			\includegraphics[width=0.6\linewidth]{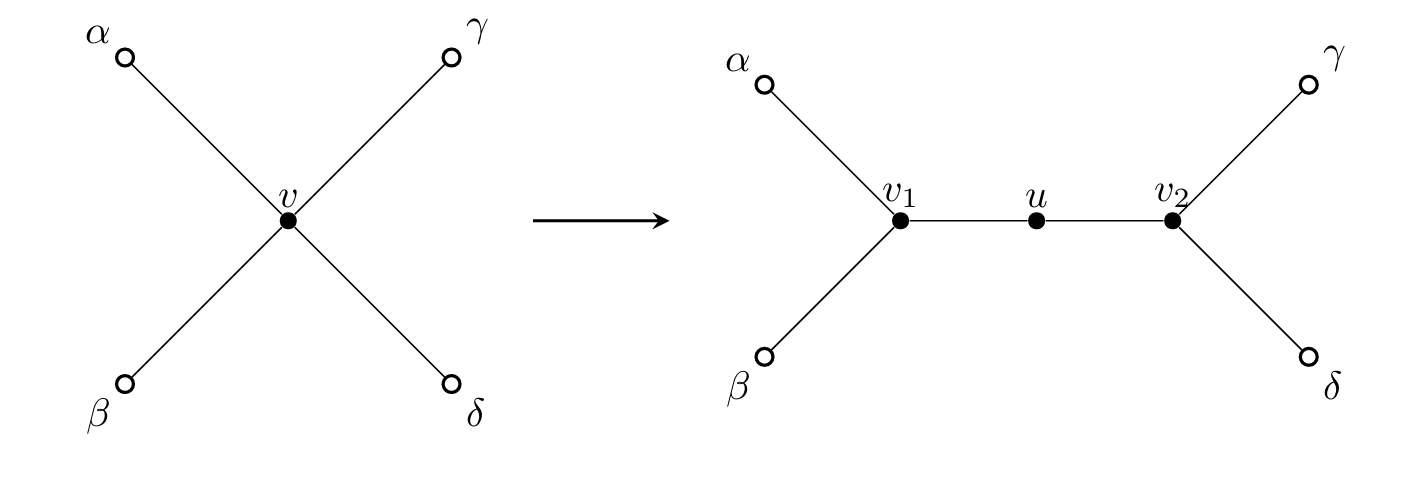}
			\caption{The transformation of a degree 4 vertex in the $K_{1,5}$- free bipartite graph}
			\label{fig:ST2}
		\end{figure}
		\textbf{Claim}: $ST-P(G,R,k)$ if and only if $ST(G',R'=R,k'=3k)$.\\
		\textit{Necessity}: Suppose $v$ is a vertex in the steiner set $S$ for $ST-P(G,R,k)$ then the steiner set $S'$ corresponding to $ST(G',R',k')$ contains the vertices $v_1,u,v_2$. So for each vertex in $S$, there will be 3 vertices in $S'$.\\ \textit{Sufficiency}: Only two possibilities for each non-terminal vertices - either all three of $\{v_1,u,v_2\}$ are in $S'$ (for each $v$) or none of the vertex are in $S'$. In first case include the corresponding $v$ in $S$ and in second case exclude the corresponding $v$ from $S$ .
		
	\end{proof}
	
	Combining all these results, we get the following dichotomy 
	\begin{definition3}\label{thm2}
		ST in a $K_{1,r}$- free bipartite graph is polynomial time solvable for $1\leq r\leq3$ and NP- complete for $r\geq4$.
	\end{definition3}
	
	Now we extend this study to other $k$-split graphs.
	ST was proved to be NP-complete on bisplit graphs \cite{inbook}. We will show that the problem is NP‑complete on trisplit graphs, and likewise for other $k$-split graphs.
	\begin{definition1}\label{frbisp}
		If $r$ is fixed , the class of $K_{1,{r}}$-free bisplit graphs is finite.
	\end{definition1}
	\begin{proof}
		Consider a $K_{1,{r}}$-free bisplit graph $G=A\cup B \cup I$, where $A\cup B$ forms the biclique. Then $|A| \leq r-1$ and $|B| \leq r-1$. If $|A|= r$, a vertex from $B$ forms a $K_{1,{r}}$ with all the vertices of $A$. Therefore $|A\cup B| \leq 2(r-1)$. Now suppose $|I|=2(r-1)^2+1$. Then by the Pigeon Hole Principle, there is a vertex $v$ in $A\cup B$ which is adjacent to $r$ vertices in $I$, which induces a $K_{1,{r}}$. So $|I|\leq 2(r-1)^2$. Thus $|V(G)| \leq 2(r-1)+2(r-1)^2=2r(r-1)$. Hence, we can conclude that the class of $K_{1,{r}}$-free bisplit graphs is finite.
	\end{proof}
	This is an interesting property because for a fixed $r$, both $K_{1,{r}}$ -free split graph and $K_{1,{r}}$ -free bipartite graph are infinite classes. 
	\begin{definition3}\label{thm3}
		If $r$ is fixed, ST is polynomial time solvable for $K_{1,{r}}$-free bisplit graphs.
	\end{definition3}
	
	Now we need to prove that the problem is NP-complete when $r$ is a variable.
	
	\noindent\fbox{%
		\parbox{\textwidth}{%
			$EXACT$-$l$-$COVER(X,C)$\\
			Instance: A collection C of $l$-element subsets of a set $X = \{x_1,x_2,...,x_{lq}\}$. \\ 
			Question: Is there a sub-collection $C' \subseteq C$ such that for every $x_i \in X$,$x_i$  belongs to exactly one member of $C'$?
		}%
	}
	
	For $l\geq 3$, often abbreviated as $XlC$, is known to be NP-complete \cite{GareyJohn}.
	\begin{definition3}\label{bisnpc}
		If $r$ is a variable, then ST is NP-Complete for $K_{1,{r}}$-free bisplit graphs.
	\end{definition3}
	\begin{proof}
		An instance of $XlC(X,C)$ is reduced to an instance of $ST(G,R,k)$ in $K_{1,{r}}$-free bisplit graphs as follows:$V(G)=A\cup B\cup I \cup I'$ where $A=\{v_i:c_i \in C\}$, $B=\{v'_i:c_i \in C\}$, $I=\{u_i:x_i \in X\}$ and $I'=\{u'_i:x_i \in X\}$. We construct a biclique on the vertex sets $A\cup B$. $E(A\cup I)=\{(v_i,u_j):c_i\in C, x_j\in X  \text{ and } x_j \in c_i \}$ and $E(B\cup I')=\{(v'_i,u'_j):c_i\in C, x_j\in X  \text{ and } x_j \in c_i \}$. Let $R=I\cup I'$.\\ 
		\textbf{$G$ is a  $K_{1,r}-$ free bisplit graph, where $r=f(l)$:}
		$G$ is bisplit with $A\cup B$ forms a biclique and $I\cup I'$ forms an independent set. A star with its centre in the independent set has maximum degree $|C|$ and a star with its centre in the biclique has maximum degree $|C|+l$.    Therefore $G$ is $K_{1,{r}}$-free, where $r=|C|+l+1$ .\\
		\textbf{Claim}:$XlC(X,C)$ if and only if $ST(G,R,k)$.\\
		\textbf{Proof}.\textit{Necessity}: If there exists $C'\subseteq C$,$|C'|=\frac{|X|}{l}$ which covers all the elements of $X$, then the set of vertices $S=\{v_i \in A,v'_i \in B |c_i \in C' \} $ forms a Steiner set in $G$ with $R=I\cup I'$, where $|S|=2\frac{|X|}{l}$. \\ \textit{Sufficiency}: If there exists a Steiner set $S\subseteq A\cup B$ in $G$ on at most $k=2\frac{|X|}{l}$ Steiner vertices, then $\forall v\in S, d^{I\cup I'}(v)=l, |S|=2\frac{|X|}{l} \text{ and } N^{A\cup B}(S)=I \cup I'$, which implies that there does not exist $u,v \in S$ such that 	$N^{A\cup B}(u)\cap N^{A\cup B}(v)\neq \emptyset$. Therefore the set $C'=\{c_i\in C|v_i \in S \}$ forms an exact-l -cover of $X$.\\
	\end{proof}
	Now we extend this study to trisplit graphs.
	\begin{definition3}\label{Th6}
		ST is NP-complete for Trisplit graphs.
	\end{definition3}
	\begin{proof}
		An instance of $X3C(X,C)$ is reduced to an instance of $ST(G,R,k)$ in trisplit graphs as follows:$V(G)=A\cup B\cup D\cup I \cup I'\cup I''$ where $A=\{v_i:c_i \in C\}$, $B=\{v'_i:c_i \in C\}$, $D=\{v''_i:c_i \in C\}$ $I=\{u_i:x_i \in X\}$ ,$I'=\{u'_i:x_i \in X\}$and $I''=\{u''_i:x_i \in X\}$. We construct a triclique on the vertex sets $A\cup B\cup D$. $E(A\cup I)=\{(v_i,u_j):c_i\in C, x_j\in X  \text{ and } x_j \in c_i \}$, $E(B\cup I')=\{(v'_i,u'_j):c_i\in C, x_j\in X  \text{ and } x_j \in c_i \}$ and $E(D\cup I'')=\{(v''_i,u''_j):c_i\in C, x_j\in X  \text{ and } x_j \in c_i \}$. $R=I\cup I'\cup I''$.\\ $G$ is a trisplit with independent set $I\cup I'\cup I''$.\\
		\textbf{Claim:}:$X3C(X,C)$ if and only if $ST(G,R,k)$.\\
		\textbf{Proof}.\textit{Necessity}: If there exists $C'\subseteq C$,$|C'|=\frac{|X|}{3}$ which covers all the elements of $X$, then the set of vertices $S=\{v_i \in A,v'_i \in B,v''_i \in D |c_i \in C' \} $ forms a Steiner set in $G$ with $R=I\cup I'\cup I''$, where $|S|=\frac{|X|}{3}$.\\ \textit{Sufficiency}: If there exists a Steiner set $S\subseteq A\cup B\cup D$ in $G$ on at most $k=|X|$ Steiner vertices, then $\forall v\in S, d^{I\cup I'\cup I''}(v)=3, |S|=|X| \text{ and } N^{A\cup B\cup D}(S)=I \cup I'\cup I''$, which implies that there does not exist $u,v \in S$ such that 	$N^{A\cup B \cup D}(u)\cap N^{A\cup B\cup D}(v)\neq \emptyset$. Therefore the set $C'=\{c_i\in C|v_i \in S \}$ forms an exact-3 -cover of $X$.
	\end{proof}
	
	We are going to explore forbidden subgraphs of Trisplit graphs. We present a result that generalises Lemma \ref{frbisp}.
	\begin{definition1}\label{A2}
		If $r$ is fixed , the class of $K_{1,{r+1}}$-free trisplit graphs is finite.
	\end{definition1}
	\begin{definition3}\label{thm10}
		If $r$ is fixed, ST is polynomial time solvable for $K_{1,{r}}$-free trisplit graphs.
	\end{definition3}
	\begin{proof}
		Consider a $K_{1,{r}}$-free trisplit graph $G=A\cup B \cup C \cup I$, where $A\cup B\cup C$ forms the triclique. Then $|A| \leq r-1$, $|B| \leq r-1$ and $|C| \leq r-1$. If $|A|= r$, a vertex from $B$ forms a $K_{1,{r}}$ with all the vertices of $A$. Therefore $|A\cup B\cup C| \leq 3(r-1)$. Now suppose $|I|=3(r-1)^2+1$. Then by Pigeon Hole Principle, there is a vertex $v$ in $A\cup B\cup C$ which is adjacent to $r$ vertices in $I$, which induces a $K_{1,{r}}$. So $|I|\leq 3(r-1)^2+1$. Thus $|V(G)| \leq 3(r-1)+3(r-1)^2=3r(r-1)$. Hence, we can conclude that the class of $K_{1,{r}}$-free trisplit graphs is finite.
	\end{proof}
	
	By a similar reduction used for bisplit in Theorem \ref{bisnpc}, we get the following hardness result. 
	\begin{definition3}
		ST is NP-complete for $K_{1,{r}}$-free Trisplit graphs for a variable $r$.
	\end{definition3}
	
	\subsection{Chordal split-like graphs}
	
	It is known that ST on chordal graphs is NP-complete \cite{white}. Hence, a natural interest is to analyse the complexity of ST on chordal split-like graphs. Since split graphs constitute a proper subclass of chordal graphs, the class of chordal split graphs coincides precisely with the class of split graphs, ST is NP-complete.
	\begin{definition3}
		ST on Chordal Bipartite graphs is NP-complete \cite{MULLER1987257}.
	\end{definition3}
	\begin{definition3}
		ST on Chordal Bisplit graphs is solvable in polynomial time \cite{inbook}.
	\end{definition3}
	\begin{definition1}\label{A3}
		A graph $G$ is a Chordal Trisplit graph if and only if the following properties are satisfied
		\begin{enumerate}
			\item The triclique in $G$ is $K_{1,1,m}$, for some $m\geq 0$.
			\item For each vertex $u \in I$, $deg(u)\leq 3$ and no $u\in I$ is adjacent to two vertices in a single partition of triclique.
		\end{enumerate}
	\end{definition1}
	\begin{proof}
		\textit{Necessity}: Let $G=A\cup B \cup C\cup I$ is a Chordal Trisplit graph with $G[A\cup B \cup C]$ forms a triclique and $I$ an independent set.
		\begin{enumerate}
			\item Suppose the triclique has two partitions with at least 2 vertices. Then $G$ has $C_4$ as an induced subgraph, which contradicts the fact that $ G$ is a chordal graph.
			\item If $deg(u)\geq 4$ for some $u\in I$ , then by the Pigeon Hole Principle $u$ contains two neighbors in exactly one of the partition $A, B\text{ or }C$, let it be $x,y \in B$. Now,both $x$ and $y$ are adjacent to an $z\in A$. Vertices $u,x,y,z$ induce a $C_4$ in $G$, which contradicts that $G$ is chordal. The same contradiction occurs even if $deg(u)=2 \text{ or } 3$, but both neighbours are in exactly one of the partitions.
		\end{enumerate}
		\textit{Sufficiency}:
		Suppose $G$ is a graph satisfying (1)  and (2). Clearly, $G$ is a trisplit graph. Let $|C|=m$ and having an induced $C_4=\{a,b,c,d\}$.\\
		Case 1: $a,c\in I$. Then $b \text{ and } d$ should be in different partitions of the triclique, by (2). This will induce a chord between $b \text{ and } d$, a contradiction.\\
		Case 2: $a,c\in C$. By (2) $b, d \notin I$. Then the only possibility is $b \in B$ and $d \in A$ (or vice-versa). But $b \text{ and } d$ are adjacent, a contradiction. 
	\end{proof}
	From this property, we derive the Algorithm \ref{Algorithm 1} to solve ST.
	\begin{algorithm}
		\caption{ST on Chordal Trisplit graphs}
		\begin{algorithmic}[1]\label{Algorithm 1}
			\REQUIRE A connected chordal trisplit graph \( G \) with \( R \subseteq I \)
			\ENSURE A Steiner solution \( S \) for \( G \)
			\STATE Let the vertices in the triclique be \( a, b, c_1\ldots, c_m \)
			\IF{there exists a pendent vertex \( v \in R \)}
			\STATE \( S \gets S \cup N(v) \)
			\STATE \( R \gets R \setminus N(N(v)) \)
			\ENDIF
			\IF{\( R = \emptyset \) \AND \( G[R \cup S] \) is connected}
			\STATE \textbf{stop the algorithm}
			\ELSE
			\FOR{\( P \text{ in } \{S\cup \{a\},S\cup \{b\},S\cup \{a,b\}\}\)}
			\IF{\(R \subseteq N(P)\)}
			\STATE \(S=P\) 
			\STATE \textbf{break}
			\ENDIF
			\ENDFOR
			\ENDIF
			\RETURN \( S \)
		\end{algorithmic}
	\end{algorithm}
	\begin{definition3}\label{Thr9}
		Finding a minimum Steiner tree on chordal trisplit graphs is solvable in polynomial time.
	\end{definition3}
	\begin{proof}
		For a pendent vertex, its only neighbour should be added to the $S$. This is ensured by steps 3 and 4. For every other vertex, at least one neighbour must be $a$ or $b$. So one of the $\{S,S\cup \{a\},S\cup \{b\},S\cup \{a,b\}\}$ is the minimum Steiner set. Since we check for the presence of pendant vertices in $G$, the check can be done in linear time. Thus the time complexity of Algorithm is $O(n)$, where $n$ is the number of vertices in $G$.
	\end{proof}
	We can extend this algorithm for chordal k-split graphs, so:
	\begin{definition3}
		Finding a minimum Steiner tree on chordal $k$-split graphs is solvable in polynomial time, $k\geq 2$ is fixed.
	\end{definition3}
	The Steiner tree problem in an arbitrary Split graph is NP-complete. Consider a split graph $G=K+I$ with $|K|\leq k$. Then, $G$ can be represented as a $k$-split graph in which each partition consists of a single vertex. Also, since a split graph is a chordal graph, $G$ is a chordal $k$-split graph. Then the following result
	\begin{definition3}
		ST is NP-complete for chordal $k$-split graphs for a variable $k$.
	\end{definition3}
	
	\subsection{ST with diameter as a parameter}
	Diameter of a split graph is at most 3, while the diameter of a bisplit graph is known to be at most 4 \cite{inbook}. But there is no such bound possible for the class of bipartite graphs. Here, we will show that the diameter of a trisplit graph is also at most 4. Then we will study the complexity of ST on such graph classes with diameter as a parameter. 
	
	The split graphs instance generated from \cite{white} are split graphs of diameter three. Therefore, for a split graph of diameter three, ST is NP-complete. Also, split graphs of diameter two have a universal vertex. So it is trivial that ST is polynomial-time solvable.
	
	The bipartite graph instance generated from Theorem \ref{Th13} is of diameter four. Therefore, for a bipartite graph of diameter four, ST is NP-complete. By simple reduction from this problem recursively, it is possible to show that ST is NP-complete for bipartite graph, if the diameter is restricted to 6,8, etc. Bipartite graphs of diameter two are nothing but a complete bipartite graph, for which ST is polynomial-time solvable. The computational complexity of ST on bipartite graphs of diameter 3,5,\dots remains unsolved.
	\begin{definition1}\label{lem5}
		A trisplit graph has a diameter of at most four.
	\end{definition1}
	\begin{proof}
		Suppose $G=A\cup B\cup C + I$ is a trisplit graph with diameter five. Then there exist two vertices $x,y \in V(G)$ such that $d(x,y)=5$. If both are from triclique, their distance is at most two. For $x\in I$ and $y\in A\cup B\cup C$, $d(x,y)\leq 3$. Therefore, the only option that remains is $x,y \in I$. Consider the path $(x,a,b,u,v,y)$ in $G$. Clearly $a,v \in A\cup B\cup C$, which contradicts the fact that $d(a,v)\leq 2$.
	\end{proof}
	\begin{definition2}
		For a bisplit graph of diameter at most four, ST is NP-complete \cite{inbook}.
	\end{definition2}
	\begin{definition2}\label{pro5}
		A graph $G=A\cup B\cup I$ is a bisplit graph of diameter at most three, if and only if for any $x,y \in I$, one of the following properties is satisfied
		\begin{itemize}
			\item $N(x)\cap N(y)\neq \phi$
			\item $N(x)\cup N(y) \not\subset A$ and  $N(x)\cup N(y)\not\subset B$
		\end{itemize}
	\end{definition2}
	\begin{proof}
		Suppose $G$ is a bisplit graph of diameter at most three. If $x,y \in I$ such that $N(x)\cap N(y)= \phi$. Therefore $d(x,y)\geq 3$. Assume that $N(x)\cup N(y) \subset A$. Since the distance between any two vertices in $A$ is two, $d(x,y)= 4$, a contradiction. Similarly in case of $B$.\\
		Conversely, assume the given two conditions. Without loss of generality, we consider the following remaining cases
		\begin{enumerate}
			\item $N(x)\subseteq A$ and $N(y)\subseteq B$: then $d(x,y)= 3$.
			\item $N(x)\cap A \neq \phi$, $N(x)\cap B \neq \phi$ and $N(y)\subseteq B$: then $d(x,y)= 3$.
			\item $N(x)\cap A \neq \phi$, $N(x)\cap B \neq \phi$ and $N(y)\cap A \neq \phi$, $N(y)\cap B \neq \phi$ : then $d(x,y)= 3$.
		\end{enumerate}
		If both $x$ and $y$ are from a biclique, their distance is at most two. For $x\in I$ and $y\in A\cup B$, $d(x,y)\leq 3$. Thus $G$ is a bisplit graph having a diameter at most three.
	\end{proof}
	
	In a bisplit graph of diameter two, the presence of a universal vertex directly implies that
	\begin{definition4}
		Let $G$ be a bisplit graph of diameter two, then ST is polynomial-time solvable.
	\end{definition4}
	\begin{definition2}\label{pro6}
		For a bisplit graph of diameter three, ST is NP-complete.
	\end{definition2}
	\begin{proof}
		An instance of $X3C(X,C)$ is reduced to an instance of $ST(G,R,k)$ in bisplit graphs as follows:$V(G)=A\cup B\cup I $ where $A=\{v_i:c_i \in C\}$, $B=\{v'_i:c_i \in C\}$ and $I=\{u_i:x_i \in X\}$. We construct a biclique on the vertex sets $A\cup B$. $E(A\cup I)=\{(v_i,u_j):c_i\in C, x_j\in X  \text{ and } x_j \in c_i \}$, $E(B\cup I)=\{(v'_i,u_j):c_i\in C, x_j\in X  \text{ and } x_j \in c_i \}$. $R=I$.\\ By Proposition \ref{pro5},  $G$ is a bisplit graph of diameter three.\\
		\textbf{Claim:}:$X3C(X,C)$ if and only if $ST(G,R,k)$.\\
		\textbf{Proof}.\textit{Necessity}: If there exists $C'\subseteq C$,$|C'|=\frac{|X|}{3}$ which covers all the elements of $X$, then the set of vertices $S'=\{v_i \in A|c_i \in C' \} $ covers $R$, but the induced graph need not be connected. So remove any one vertex $v_j$ from $S'$ and replaced with corresponding $v'_j$, forms a Steiner set $S$ in $G$ for $R=I$, where $|S|=\frac{|X|}{3}$. \\ \textit{Sufficiency}: If there exists a Steiner set $S\subseteq A\cup B$ in $G$ on at most $k=\frac{|X|}{3}$ steiner vertices, then $\forall v\in S, d^{I}(v)=3, |S|=|X| \text{ and } N^{A\cup B}(S)=I$, which implies that there does not exist $u,v \in S$ such that 	$N^{A\cup B }(u)\cap N^{A\cup B}(v)\neq \emptyset$. Therefore the set $C'=\{c_i\in C|v_i \in S \}$ forms an exact-3 -cover of $X$.\\
	\end{proof}
	
	Since the trisplit graphs instance generated from Theorem \ref{Th6} are trisplit graphs of diameter at most four, the following result:
	\begin{definition2}\label{protri1}
		For a trisplit graph of diameter four, ST is NP-complete.
	\end{definition2}
	
	Proposition \ref{pro5} admits an extension to trisplit graphs, and through the analogous reduction employed in Proposition \ref{pro6}, we can show that
	\begin{definition2}\label{protri}
		For a trisplit graph of diameter three, ST is NP-complete.
	\end{definition2}
	\subsection{Star convex split-like graphs}
	
	If we introduce convex ordering on one of the partitions of a split graph, ST is polynomial-time solvable for tree-convex split graphs with convexity on the clique, whereas ST is NP-complete on tree-convex split graphs with convexity on the independent set \cite{JOUR}. Here, we are going to introduce convex ordering on one of the partitions of a bipartite graph- Star convexity. ST on the Triad convex bipartite graph and circular convex bipartite graph are still open.
	
	\begin{definition3}\label{pro7}
		A bipartite graph $G=X\cup Y$ is a star convex bipartite graph if and only if there exists a vertex $x\in X$ such that every $y\in Y$ is either a pendant vertex or is adjacent to $x$ \cite{PANDEY201951}.
	\end{definition3}
	
	\begin{definition3}\label{Th13}
		For a star convex bipartite graph, ST is NP-complete. 
	\end{definition3}
	\begin{proof}
		An instance of $X3C(X,C)$ is reduced to an instance of $ST(G,R,k)$ in star convex bipartite as follows: $V(G)=A\cup B\cup\{v\} $ where $A=\{v_i:c_i \in C\}$ and $B=\{u_i:x_i \in X\}$. $E_1=\{(v_i,u_j):c_i\in C, x_j\in X  \text{ and } x_j \in c_i \}$ and $E_2=\{(v,v_i):c_i\in C\}$. $R=B\cup \{v\}$.\\ $G$ is a bipartite graph with partitions $Z=B\cup \{v\}$ and $Y=A$. Since every vertex in $Y$ is adjacent to the vertex $v$ in $Z$, by Proposition \ref{pro7}, $G$ is a star convex bipartite graph. \\
		\textbf{Claim}: $X3C(X,C)$ if and only if $ST(G,R,k)$.\\
		\textbf{Proof}.\textit{Necessity}: If there exists $C'\subseteq C$,$|C'|=\frac{|X|}{3}$ which covers all the elements of $X$, then the set of vertices $S=\{v_i \in A|c_i \in C' \} $ covers $R$. Since the induced graph is connected $S$ forms a steiner set $G$ for $R=B$, where $|S|=\frac{|X|}{3}$. \\ \textit{Sufficiency}: If there exists a steiner set $S\subset A$ in $G$ on at most $k$ steiner vertices, then $\forall v\in S, d^{B}(v)=3 (\text{since } d^{Z}(v)=4), |S|=\frac{|X|}{3} \text{ and } N^{B}(S)=B (\text{since } N^{Z}(S)=R)$, which implies that there does not exist $u,v \in S$ such that 	$N^{ B }(u)\cap N^{ B}(v)\neq \emptyset$. Therefore the set $C'=\{c_i\in C|v_i \in S \}$ forms an exact-3 -cover of $X$.\\
	\end{proof}
	
	An analogous result to Theorem \ref{pro7} holds for bisplit graphs (convexity on biclique and independent side), and more generally for k‑split graphs.
	\begin{definition2}\label{bisstar}
		A bisplit graph $G=(A\cup B)+ I$ is a star convex bisplit graph (convexity on biclique) if and only if there exists a vertex $x\in A\cup B$ such that every $y\in I$ is either a pendant vertex or is adjacent to $x$.
	\end{definition2}
	
	Consider the Steiner Tree problem on star‑convex bisplit graphs, with convexity defined on the biclique. By the preceding result, 
	for any terminal set $T$ contained in the independent side (all remaining cases can be reduced to this one), there exists a vertex $x$ in the biclique such that every non‑pendant vertex is adjacent to $x$. So the initial Steiner set $S'$ for the given $T$ can be formed by collecting all neighbours of pendant vertices, and for non-pendent vertices, add $x$ to the set. But the induced graph $G[T\cup S']$ may not be connected. This is because $S'$ lies entirely within one partition of the biclique. So adding any vertex from the other partition to $S'$ solves the problem. All this procedure can be done in polynomial time.
	\begin{definition3}\label{T21}
		ST on star‑convex bisplit graphs, with convexity defined on the biclique, is polynomial-time solvable.
	\end{definition3}
	\begin{definition3}\label{A4}
		ST on star‑convex bisplit graphs, with convexity defined on the independent side, is NP-complete.
	\end{definition3}
	\begin{proof}
		An instance of $X3C(X, C)$ is reduced to an instance of $ST(G,R,k)$ in star‑convex bisplit graphs, with convexity defined on the independent side, as follows:$V(G)=A\cup B\cup I' \cup I'' \cup \{u\}$ where $A=\{v_i:c_i \in C\}$, $B=\{v'_i:c_i \in C\}$, $I'=\{u_i:x_i \in X\}$ and $I''=\{u'_i:x_i \in X\}$. We construct a biclique on the vertex sets $A\cup B$. $E(A\cup I')=\{(v_i,u_j):c_i\in C, x_j\in X  \text{ and } x_j \in c_i \}$ and $E(B\cup I'')=\{(v'_i,u'_j):c_i\in C, x_j\in X  \text{ and } x_j \in c_i \}$. $u$ is made universal to all vertices in $A\cup B$. Let $R=I\cup I'\cup \{u\}$.\\ 
		\textbf{$G$ is a star‑convex bisplit graph, with convexity defined on the independent side,:}
		$G$ is bisplit with $A\cup B$ forms a biclique and $I'\cup I''\cup \{u\}$ forms an independent set. Since every vertex of $A\cup B$ is adjacent to $u$, by Proposition \ref{bisstar}, $G$ is a star‑convex bisplit graph, with convexity defined on the independent side. \\
		\textbf{Claim:}$X3C(X,C)$ if and only if $ST(G,R,k)$.\\
		\textbf{Proof}.\textit{Necessity}: If there exists $C'\subseteq C$,$|C'|=\frac{|X|}{3}$ which covers all the elements of $X$, then the set of vertices $S=\{v_i \in A,v'_i \in B |c_i \in C' \} $ forms a Steiner set in $G$ with $R=I'\cup I''\cup\{u\}$, where $|S|=2\frac{|X|}{3}$. \\ \textit{Sufficiency}: If there exists a Steiner set $S\subseteq A\cup B$ in $G$ on at most $k=2\frac{|X|}{3}$ Steiner vertices, then $\forall v\in S, d^{I'\cup I''}(v)=3, |S|=2\frac{|X|}{3} \text{ and } N^{A\cup B}(S)=R$, which implies that there does not exist $u,v \in S$ such that 	$N^{A\cup B}(u)\cap N^{A\cup B}(v)\neq \emptyset$. Therefore the set $C'=\{c_i\in C|v_i \in S \}$ forms an exact-3 -cover of $X$.\\
	\end{proof}
	
	Both results can be extended naturally to k-split graphs, $k\geq 3$.
	\section{ST in $K_{1,r}$-free Chordal graphs}
	In Splits graphs, a notable subclass of Chordal graphs, computational hardness starts from $K_{1,5}$-free graphs. Consequently, ST is NP-complete even within   $K_{1,5}$-free chordal graphs. Furthermore, undirected path graphs, which form another strict subclass of chordal graphs, also exhibit  NP-completeness for ST \cite{DEFIGUEIREDO2022184}. By the prescribed construction, the graph obtained by reduction is $K_{1,4}$-free. This will help us to update the dichotomy status of $K_{1,r}$-free chordal graphs.
	\begin{definition3}\label{thm5}
		ST on $K_{1,4}$- free undirected path graphs is NP-complete.
	\end{definition3}
	
	\begin{proof}
		Let \( P = \{p_1, \ldots, p_n\} \), \( Q = \{q_1, \ldots, q_n\} \), and \( R = \{r_1, \ldots, r_n\} \) be disjoint sets, each of cardinality \( n \), for some positive integer \( n \). Let \( S = \{s_1, \ldots, s_m\} \subseteq P \times Q \times R \) be a subset of cardinality \( m \), for some positive integer \( m \).
		Define the instance of 3D-Matching as \( I = (P, Q, R, S) \), constituted by the sets \( P, Q, R \), and \( S \).
		Construct a graph \( G \) from the instance \( I \) as follows:  \\
		For each \( s_j \in S \), define the set
		\[
		V_j = \{a_j, b_j, c_j, x_j, y_j, z_j^1, z_j^2, z_j^3\}.
		\]
		Define the vertex set of the graph \( G \) as
		\[
		V(G) = \bigcup_{j=1}^m V_j \cup P \cup Q \cup R.
		\]
		
		Let
		\[
		K = \bigcup_{s_j \in S} \{a_j, b_j, c_j, x_j\}
		\]
		be a clique in \( G \).
		
		For each \( s_j \in S \), the following subsets are cliques in \( G \):
		\[
		\{a_j, b_j, x_j, y_j\}, \quad \{a_j, y_j, z_j^1\}, \quad \{b_j, y_j, z_j^2\}, \quad \{c_j, x_j, z_j^3\}.
		\]
		
		Additionally, for each element:\\
		- \( p_i \in P \), the set \( \{p_i\} \cup \{a_j : p_i \in s_j, s_j \in S\} \) is a clique in \( G \);\\
		- \( q_i \in Q \), the set \( \{q_i\} \cup \{b_j : q_i \in s_j, s_j \in S\} \) is a clique in \( G \);\\
		- \( r_i \in R \), the set \( \{r_i\} \cup \{c_j : r_i \in s_j, s_j \in S\} \) is a clique in \( G \).\\

		It is enough to show that the graph $G$ obtained by the reduction in \cite{DEFIGUEIREDO2022184} is $K_{1,4}$-free.
		\begin{enumerate}
			\item A star with $p_i$ as center: Maximum induced star is $K_{1,1}$, because it's only neighbors $a_j$ forms a clique in $G$. Similarly, for a star with $q_i$ or $r_i$ as the centre.
			\item A star with $a_j$ as centre: Since it's a part of the clique $K$, at most one neighbour can be taken from that clique set. Each $a_j$ is adjacent to exactly one $p_i$, which is non adjacent to $\cup \{b_j,c_j,x_j\}$. Now take one neighbor from each set $\{b_j,x_j,y_j\}$ and $\{z^1_j,y_j\}$. Maximum size independent sets are of size 2: $\{b_j,z^1_j\}\text{ and }\{x_j,z^1_j\}$. Along with $p_i$ it induces a $K_{1,3}$. So the maximum induced star is $K_{1,3}$. Similarly, for a star with $b_j$ or $c_j$ as the centre.
			\item A star with $x_j$ as centre:  Since it's a part of the clique $K$, at most one neighbour can be taken from that clique set. Now take one neighbor from each set $\{a_j,b_j,y_j\}$ and $\{z^3_j,c_j\}$. Maximum size independent sets are of size 2: are $\{a_j,z^3_j\},\{y_j,c_j\},\{y_j,z^3_j\}\text{ and }\{b_j,z^3_j\}$, which forms a $K_{1,2}$. Only for the set $\{y_j,z^3_j\}$ along with a vertex from $K$, which is non adjacent to $y_j$ (i.e. $y_k $ such that $k\neq j$) forms a $K_{1,3}$. So the maximum induced star is $K_{1,3}$.
			\item A star with $y_j$ as center: Take one neighbor from each set $\{a_j,b_j,x_j\}$ $\{a_j,z^1_j\}$ and $\{z^2_j,b_j\}$. Maximum size independent sets are of size 3:  $\{a_j,z^1_j,z^2_j\}$ and $\{x_j,z^1_j,z^2_j\}$. So the maximum induced star is $K_{1,3}$.
			\item A star with $z^1_j$ as centre: Maximum induced star is $K_{1,1}$, because it's a part of a single triangle only. Similarly, for a star with $z^2_j$ or $z^3_j$ as the centre.
		\end{enumerate}
		
		So in all possible cases the maximum induced star in $G$ is $K_{1,3}$.
	\end{proof}
	\begin{definition4}\label{thm6}
		ST on $K_{1,4}$- free chordal graphs is NP-complete.
	\end{definition4}

	\section{Conclusion}
	In this paper, we studied the Steiner tree problem on split-like graphs under the structural constraints - star-free, bounded diameter, chordality and star convexity, which provide several dichotomy results. Beyond star convexity, the problem can also be studied under other convexity frameworks such as triad convexity, tree convexity, comb convexity, etc. All NP-completeness results extend naturally to the weighted version of the Steiner tree problem, whereas the polynomial-time solvable cases must be analyzed separately. The Steiner tree problem remains open for diameter-three bipartite graphs and claw-free chordal graphs. Also we can extend this study to other variants of Steiner tree problem.  Furthermore, fixed-parameter
	tractability (FPT) and approximation algorithms for different graph classes provide promising direction.
	
	\bibliography{splitlikbib} 

@article{garey,
	title={A Guide to the Theory of NP-Completeness},
	author={Garey, Michael R and Johnson, David S},
	journal={Computers and intractability},
	pages={641--650},
	year={1979}
}

@Inbook{Karp1972,
	author="Karp, Richard M.",
	editor="Miller, Raymond E.
	and Thatcher, James W.
	and Bohlinger, Jean D.",
	title="Reducibility among Combinatorial Problems",
	bookTitle="Complexity of Computer Computations: Proceedings of a symposium on the Complexity of Computer Computations, held March 20--22, 1972, at the IBM Thomas J. Watson Research Center, Yorktown Heights, New York, and sponsored by the Office of Naval Research, Mathematics Program, IBM World Trade Corporation, and the IBM Research Mathematical Sciences Department",
	year="1972",
	publisher="Springer US",
	address="Boston, MA",
	pages="85--103",
	isbn="978-1-4684-2001-2",
	doi="10.1007/978-1-4684-2001-2_9",
	url="https://doi.org/10.1007/978-1-4684-2001-2_9"
}

@article{RENJITH2020246,
	title = {The Steiner tree in K1,r-free split graphs—A Dichotomy},
	journal = {Discrete Applied Mathematics},
	volume = {280},
	pages = {246-255},
	year = {2020},
	note = {Algorithms and Discrete Applied Mathematics (CALDAM 2016)},
	issn = {0166-218X},
	doi = {https://doi.org/10.1016/j.dam.2018.05.050},
	url = {https://www.sciencedirect.com/science/article/pii/S0166218X18303111},
	author = {P. Renjith and N. Sadagopan}
}

@article{GareyJohn,
	ISSN = {00361399},
	URL = {http://www.jstor.org/stable/2100192},
	author = {M. R. Garey and D. S. Johnson},
	journal = {SIAM Journal on Applied Mathematics},
	number = {4},
	pages = {826--834},
	publisher = {Society for Industrial and Applied Mathematics},
	title = {The Rectilinear Steiner Tree Problem is NP-Complete},
	urldate = {2025-06-23},
	volume = {32},
	year = {1977}
}

@article{white,
	author = {White, Kevin and Farber, Martin and Pulleyblank, William},
	title = {Steiner trees, connected domination and strongly chordal graphs},
	journal = {Networks},
	volume = {15},
	number = {1},
	pages = {109-124},
	doi = {https://doi.org/10.1002/net.3230150109},
	url = {https://onlinelibrary.wiley.com/doi/abs/10.1002/net.3230150109},
	eprint = {https://onlinelibrary.wiley.com/doi/pdf/10.1002/net.3230150109},
	year = {1985}
}

@article{inbook,
	title = {Bisplit graphs — A structural and algorithmic study},
	journal = {Discrete Applied Mathematics},
	volume = {389},
	pages = {243-253},
	year = {2026},
	issn = {0166-218X},
	doi = {https://doi.org/10.1016/j.dam.2026.04.006},
	url = {https://www.sciencedirect.com/science/article/pii/S0166218X2600209X},
	author = {A. Mohanapriya and P. Renjith and N. Sadagopan}
}

@article{DEFIGUEIREDO2022184,
	title = {Revising Johnson’s table for the 21st century},
	journal = {Discrete Applied Mathematics},
	volume = {323},
	pages = {184-200},
	year = {2022},
	note = {LAGOS’19 – X Latin and American Algorithms, Graphs, and Optimization Symposium – Belo Horizonte, Minas Gerais, Brazil},
	issn = {0166-218X},
	doi = {https://doi.org/10.1016/j.dam.2021.05.021},
	url = {https://www.sciencedirect.com/science/article/pii/S0166218X21002109},
	author = {Celina M.H. {de Figueiredo} and Alexsander A. {de Melo} and Diana Sasaki and Ana Silva}
}

@article{PANDEY201951,
	title = {Domination in some subclasses of bipartite graphs},
	journal = {Discrete Applied Mathematics},
	volume = {252},
	pages = {51-66},
	year = {2019},
	note = {13th Cologne-Twente Workshop on Graphs and Combinatorial Optimization (CTW 2015)},
	issn = {0166-218X},
	doi = {https://doi.org/10.1016/j.dam.2018.03.029},
	url = {https://www.sciencedirect.com/science/article/pii/S0166218X18301203},
	author = {Arti Pandey and B.S. Panda}
}

@article{article3,
	author = {Bodlaender, Hans and Brettell, Nick and Johnson, Matthew and Paesani, Giacomo and Paulusma, Daniël and Leeuwen, Erik},
	year = {2021},
	month = {03},
	pages = {},
	title = {Steiner Trees for Hereditary Graph Classes: a Treewidth Perspective},
	volume = {867},
	journal = {Theoretical Computer Science},
	doi = {10.1016/j.tcs.2021.03.012}
}

@article{MULLER1987257,
	title = {The NP-completeness of steiner tree and dominating set for chordal bipartite graphs},
	journal = {Theoretical Computer Science},
	volume = {53},
	number = {2},
	pages = {257-265},
	year = {1987},
	issn = {0304-3975},
	doi = {https://doi.org/10.1016/0304-3975(87)90067-3},
	url = {https://www.sciencedirect.com/science/article/pii/0304397587900673},
	author = {Haiko Müller and Andreas Brandstädt}
}

@article{JOUR,
	title={On convexity in split graphs: complexity of Steiner tree and domination},
	author={Mohanapriya A. , Renjith P., Sadagopan N.},
	journal={Journal of Combinatorial Optimization},
	pages={},
	year={2024}
}

@article{Seriespar,
	title = "Steiner trees, partial 2–trees, and minimum IFI networks",
	author = "Wald, \{Joseph A.\} and Colbourn, \{Charles J.\}",
	year = "1983",
	doi = "10.1002/net.3230130202",
	language = "English (US)",
	volume = "13",
	pages = "159--167",
	journal = "Networks",
	issn = "0028-3045",
	publisher = "Wiley-Liss Inc.",
	number = "2",
}

@article{COLBOURN1990179,
	title = {Permutation graphs: Connected domination and Steiner trees},
	journal = {Discrete Mathematics},
	volume = {86},
	number = {1},
	pages = {179-189},
	year = {1990},
	issn = {0012-365X},
	doi = {https://doi.org/10.1016/0012-365X(90)90359-P},
	url = {https://www.sciencedirect.com/science/article/pii/0012365X9090359P},
	author = {Charles J. Colbourn and Lorna K. Stewart}
}
	\bibliographystyle{ieeetr}
	
\end{document}